\documentclass[onecolumn,iop,usenatbib,tighten,revtex-4]{openjournal}
\usepackage[T1]{fontenc}
\usepackage{ae,aecompl}
\usepackage{pgfplots}
\usepgfplotslibrary{patchplots}
\usetikzlibrary{patterns, positioning, arrows}
\pgfplotsset{compat=1.15}
\usepackage{float}
\usepackage{adjustbox}
\usepackage{amsmath}
\usepackage{enumerate}
\usepackage{microtype}
\usepackage{orcidlink} 

\DeclareRobustCommand{\VAN}[3]{#2}
\let\VANthebibliography\thebibliography
\def\thebibliography{\DeclareRobustCommand{\VAN}[3]{##3}\VANthebibliography}

\usepackage[bitstream-charter]{mathdesign}
\usepackage{graphicx}	

\usepackage{soul}
\usepackage{microtype}
\usepackage[title]{appendix}
\usepackage{hyperref}	
\hypersetup{colorlinks=true,linkcolor=blue,citecolor=blue,filecolor=blue,urlcolor=blue}

\newcommand{\ltsima}{$\; \buildrel < \over \sim \;$}
\newcommand{\lsim}{\lower.5ex\hbox{\ltsima}}
\newcommand{\gtsima}{$\; \buildrel > \over \sim \;$}
\newcommand{\gsim}{\lower.5ex\hbox{\gtsima}}
\newcommand{\bra}{\langle}
\newcommand{\ket}{\rangle}

\newcommand{\dd}{\mathrm{d}}

\newcommand{\thetap}{\theta^\prime}
\newcommand{\likeli}{\mathcal{L}}

\DeclareSymbolFont{usualmathcal}{OMS}{cmsy}{m}{n}
\DeclareSymbolFontAlphabet{\mathcal}{usualmathcal}

\begin{document}

\title{Bayesian distances for quantifying tensions in cosmological inference and the surprise statistic\vspace{-30pt}}

\author{Benedikt Schosser\,\orcidlink{0009-0007-8905-7749}$^{1,\star}$}
\author{Pedro Riba Mello\,\orcidlink{0009-0005-0325-2400}$^{2}$}
\author{Miguel Quartin\,\orcidlink{0000-0001-5853-6164}$^{2,3,4}$}
\author{Bj{\"o}rn Malte Sch{\"a}fer\,\orcidlink{0000-0002-9453-5772}$^{1, \sharp}$}
\thanks{$^\star$ \href{mailto:schosser@stud.uni-heidelberg.de}{schosser@stud.uni-heidelberg.de}}
\thanks{$^\sharp$ \href{mailto:bjoern.malte.schaefer@uni-heidelberg.de}{bjoern.malte.schaefer@uni-heidelberg.de}}

\affiliation{$^{1}$Zentrum f{\"u}r Astronomie der Universit{\"a}t Heidelberg, Astronomisches Rechen-Institut, Philosophenweg 12, 69120 Heidelberg, Germany}
\affiliation{$^{2}$Instituto de Física, Universidade Federal do Rio de Janeiro, 21941-972, Rio de Janeiro, RJ, Brazil}
\affiliation{$^{3}$Observatório do Valongo, Universidade Federal do Rio de Janeiro, 20080-090, Rio de Janeiro, RJ, Brazil}
\affiliation{$^{4}$PPGCosmo, Universidade Federal do Espírito Santo, 29075-910, Vitória, ES, Brazil}

\begin{abstract}
Tensions between cosmological parameters derived through different channels can be a genuine signature of new physics that $\Lambda$CDM as the standard model is not able to reproduce, in particular in the missing consistency between parameter estimates from measurements the early and late Universe. Or, they could be caused by yet to be understood systematics in the measurements as a more mundane explanation. Commonly, cosmological tensions are stated in terms of mismatches of the posterior parameter distributions, often assuming Gaussian statistics. More importantly, though, would be a quantification if two data sets are consistent to each other before combining them into a joint measurement, ideally isolating hints at individual data points that have a strong influence in generating the tension. For this purpose, we start with statistical divergences applied to posterior distributions following from different data sets and develop the theory of a Fisher metric between two data sets, in analogy to the Fisher metric for different parameter choices. As a topical example, we consider the tension in the Hubble-Lema{\^i}tre constant $H_0$ from supernova and measurements of the cosmic microwave background, derive a ranking of data points in order of their influence on the tension on $H_0$. For this particular example, we compute Bayesian distance measures and show that in the light of CMB data, supernovae are commonly too bright, whereas the low-$\ell$ CMB spectrum is too high, in agreement with intuition about the parameter sensitivity.
\end{abstract}


\section{introduction}
The Bayes' theorem as the central tool of statistical inference, describes the process of updating the knowledge on the model parameters $\theta^\mu$ given a data set $y^i$ within a preselected model \citep[for summaries and applications to cosmology, see][]{fisher_logic_1935, 2006AIPC..872...31C, trotta_bayesian_2017, loredo_bayesian_2012},
\begin{equation}
p(\theta|y) = \frac{\likeli(y|\theta)\pi(\theta)}{p(y)}\, ,
\label{eq:bayes_theorem}
\end{equation}
by setting the posterior distribution $p(\theta|y)$ into relation with the likelihood $\likeli(y|\theta)$. The normalisation is given by the Bayes-evidence $p(y)$
\begin{equation}
p(y) = \int\dd^n\theta\:\likeli(y|\theta)\pi(\theta) \,,
\label{eq:evidence}
\end{equation}
where the prior distribution $\pi(\theta)$ encapsulates the knowledge on the parameters $\theta^\mu$ before the data set $y^i$ has been recorded. The likelihood $\likeli(y|\theta)$ as a conditional probability describes the distribution of data points $y$ that can be expected under a particular parameter choice $\theta$ and contains the knowledge of the physical model and the error process in the measurement. As two parameter choices $\theta$ and $\theta^\prime = \theta+\dd\theta$ will differ in their prediction of the data distribution it makes sense to quantify this difference in terms of e.g. the Kullback-Leibler divergence $\Delta S(\theta,\theta+\dd\theta) $ \citep{kullback1951,Renyi_original},
\begin{equation}
S(\theta,\theta+\dd\theta) =
\int\dd y\:\likeli(y|\theta)\ln\frac{\likeli(y|\theta)}{\likeli(y|\theta+\dd\theta)} \simeq
\frac{1}{2}F_{\mu\nu}(\theta)\dd\theta^\mu\dd\theta^\nu \,,
\label{eqn_parameter_fisher}
\end{equation}
which can be expanded with a quadratic, leading-order term, defining the Fisher metric $F_{\mu\nu}(\theta)$ \citep{tegmark_karhunen-loeve_1997, elsner_fast_2012,coe_fisher_2009,2011IJMPD..20.2559B,wolz_validity_2012, schafer_describing_2016,sellentin_breaking_2014} as a positive definite distance measure on the manifold of parameters \citep{amari_information_2016}: Effectively, it relates the question of distinguishability of the data distributions originating from two different parameter choices to a distance measure between two infinitesimally different parameter choices, which in the Gaussian case corresponds exactly to the $\Delta\chi^2$-functional.

Bayes' theorem is perfectly symmetric, as it arises from decomposing the joint distribution $p(y,\theta)$ either in terms of  $p(y,\theta) = \likeli(y|\theta)\pi(\theta)$ or equivalently, according to $p(\theta,y) = p(\theta|y)p(y)$. Therefore, it might be sensible to ask whether one can define a second Fisher metric for the posterior distribution $p(\theta|y)$ in analogy to the Fisher metric for the likelihood $\likeli(y|\theta)$, as both quantities are conditional probabilities. It would be used to quantify if an infinitesimally different data set $y^i+\dd y^i$ would have given rise to a different posterior distribution $p(\theta|y)$ according to \citep{amari_information_2016}
\begin{equation}
S(y,y+\dd y) =
\int\dd\theta\: p(\theta|y) \ln\frac{p(\theta|y)}{p(\theta|y+\dd y)} \simeq
\frac{1}{2}F_{ij}(y)\dd y^i\dd y^j,
\label{eqn_data_fisher}
\end{equation}
where $F_{ij}$ would likewise be a distance measure, but in this case in data space. It would quantify how much a second data set would differ in its induced posterior distribution. This point motivates our paper: We would like to quantify the amount of dissimilarity between two data sets, as a measure of tension. Consistency in the data sets would mean that they generate posteriors that are compatible with each other and have a small, or vanishing in the case of equal posteriors, Kullback-Leibler divergence $S(y,y+\dd y)$.

The question of consistency \citep{raveri_concordance_2019, raveri_quantifying_2020, 2017PhRvD..95l3535C, 2017PhRvD..96b3532L, 2016PhRvD..93j3507S} of parameter estimates is commonplace in post-Planck cosmology. Different observational channels suggest estimates for cosmological parameters that are in tension with each other, most notably in the Hubble-Lema{\^i}tre constant $H_0$ from the CMB and low-redshift measurements like supernovae, or the determination of $S_8$ from the CMB in comparison to weak lensing \citep{2017MNRAS.471.1259J}. The overarching physical questions in this context concern systematics, possibly of astrophysical origin, and the consistency of $\Lambda$CDM as the standard model of cosmology between the early and late Universe \citep{efstathiou_problems_2017, santos_bayesian_2016, ange_improving_2023, douspis_tension_2019, couchot_relieving_2015, 2015MNRAS.451.2877M}, and the emergence of new physics in relation to e.g. dark energy or modified gravity \citep{scott_standard_2018, babichev_introduction_2013, bernal_parameter_2015, dossett_constraints_2015, renk_galileon_2017, 2017PhRvL.119j1301H}. Given the deviation of the posterior parameter distributions from their idealised Gaussian shape and the high dimensionality of the parameter space, it is clear that Gaussian measures like the Mahalanobis distance are not fully capturing all dissimilarities between the distributions and that tools from information theory are needed \citep{nicola_consistency_2018, grandis_information_2016, handley_quantifying_2019, 10.1214/aoms/1177728069, 2014PhRvD..90b3533S}.

We review the concept of metric distances on statistical manifolds in Sect.~\ref{sect_theory} before introducing the idea of a Fisher-like metric for measuring distances between data sets in Sect.~\ref{sect_fisher}. We apply this method to the well-known tension between estimates of the Hubble-Lema{\^i}tre constant $H_0$ from supernova and cosmic microwave background data in Sect.~\ref{sect_cosmo} and summarise our main results in Sect.~\ref{sect_summary}. Throughout the paper, we work with a spatially flat $\Lambda$CDM cosmology constrained by supernova data \citep{Suzuki_2012} and by Planck's measurements of the CMB \citep{Planck2018}, with best-fit values from these two data sets. Notation-wise, we use Latin letters and indices for data tuples, Greek letters with Greek indices for parameter tuples, and impose the Einstein summation convention. Then, the parameter covariance is given as $\bra\theta^\mu\theta^\nu\ket-\bra\theta^\mu\ket\bra\theta^\nu\ket$, and for a Gaussian distribution equal to the inverse Fisher metric $F^{\mu\nu}$. It fulfills $F^{\alpha\mu}F_{\alpha\nu} = \delta^\mu_\nu$ with the Fisher metric $F_{\mu\nu}$. Derivatives $\partial_\mu = \partial/\partial\theta^\mu$ with respect to the parameters lead to linear forms, $F_{\mu\nu} = \bra\partial_\mu\ln\likeli\:\partial_\nu\ln\likeli\ket$, such that a quantity like $\Delta\chi^2 = F_{\mu\nu}\Delta\theta^\mu\Delta\theta^\nu$ is an invariant. We would like to emphasise that the $H_0$-tension is a topical and relevant example for testing our methods \citep[c.f.][]{2013PDU.....2..166V}, but we do \emph{not} intend to provide physical mechanisms or identify a systematic to resolve the tension.

\section{From entropies and the Bayes' theorem to the Fisher metric}\label{sect_theory}

\subsection{Fisher metric and relative entropies}
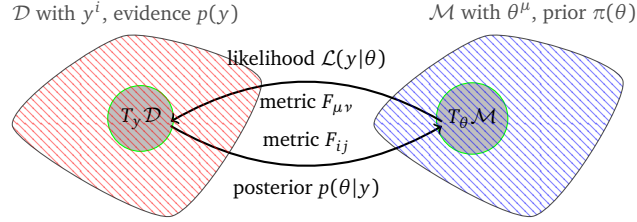
\begin{figure}
\begin{center}
\begin{tikzpicture}[scale=\textwidth/30cm]

\node[circle,draw=green,text=black,fill=lightgray,yshift=1mm,xshift=-4mm] (c) at (3,0){$T_y\mathcal{D}$};
\node[circle,draw=green,text=black,fill=lightgray,yshift=1mm,xshift=4mm] (c) at (9,0){$T_\theta\mathcal{M}$};

    \path[<-,thick] (3, 0.1) edge [bend left] node[yshift=3mm] {likelihood $\likeli(y|\theta)$} node[yshift=-3mm] {metric $F_{\mu\nu}$}(9, 0.1);

    \path[->,thick] (3, 0) edge [bend right] node[yshift=-3mm] {posterior $p(\theta|y)$} node[yshift=3mm] {metric $F_{ij}$} (9, 0);



    \draw[smooth cycle, tension=0.4, fill=white, pattern color=red, pattern=north west lines, opacity=0.7] plot coordinates{(2,2) (-0.5,0) (3,-2) (5,1)} node at (2,2.5) {$\mathcal{D}$ with $y^i$, evidence $p(y)$};

    \draw[smooth cycle, tension=0.4, fill=white, pattern color=blue, pattern=north west lines, opacity=0.7] plot coordinates{(10,2) ( 7.5,0) (11,-2) (13,1)} node at (11,2.5) {$\mathcal{M}$ with $\theta^\mu$, prior $\pi(\theta)$};
\end{tikzpicture}
\label{fig:manifolds}
\caption{Sketch of data manifold $\mathcal{D}$ and the parameter manifold $\mathcal{M}$ with the corresponding probabilities. The two conditional distributions $\mathcal{L}(y|\theta)$ and $p(\theta|y)$ imply the existence of a metric $F_{\mu\nu}(\theta)$ on the tangent space $T_\theta\mathcal{M}$ of parameters $\theta^\mu$ and of a second metric $F_{ij}(y)$ on the tangent space $T_y\mathcal{D}$ of data points $y^i$.}
\end{center}
\end{figure}

The inference process, as determined by Bayes' theorem, suggests the distribution $p(y)$ in data space $\mathcal{D}$ and the prior distribution $\pi(\theta)$ in parameter space $\mathcal{M}$. These spaces, to take up the foundational idea of information geometry, are actually manifolds. Mediating between the two manifolds are the two conditional probabilities: The likelihood $\likeli(y|\theta)$ maps from $\theta$ to $y$ as it provides a distribution of data $y$ for each choice of the parameters $\theta$, and the posterior distribution $p(\theta|y)$ as a map from $y$ to $\theta$, which describes the uncertainty on the parameters $\theta$ if the data $y$ has been taken into account, all illustrated by Fig.~\ref{fig:manifolds}.

By definition, one can assign a metric to each conditional probability, and conventionally this is done for the likelihood $\mathcal{L}(y|\theta)$ as outlined by Eq.~(\ref{eqn_parameter_fisher}). In parallel, there is a second metric associated with the posterior distribution $p(\theta|y)$ giving rise to the data space Fisher metric $F_{ij}$, as summarised by Eq.~(\ref{eqn_data_fisher}). In both cases, one would proceed by considering the difference in the distributions for an infinitesimal change in the condition, $\dd\theta$ or $\dd y$, respectively. The difference in distribution is measured by an $f$-divergence, most commonly by the Kullback-Leibler divergence. For our particular application, this choice becomes a necessity as only the Kullback-Leibler entropy measure is compatible with Bayes' law: For instance, R{\'e}nyi- or Amari-divergences \citep{amari_information_2016} would provide a sensible construction of a metric in the spirit of Eq.~(\ref{eqn_parameter_fisher}), but they neither allow an additive definition of conditional entropies nor an entropy-version of Bayes' law, $S(y|\theta) + S(\theta) = S(\theta|y) + S(y)$.

In fact, employing Bayes' law~(\ref{eq:bayes_theorem}) in conjunction with the relative entropy~(\ref{eqn_data_fisher}) between the posteriors $p(\theta|y^\prime)$ and $p(\theta|y)$ originating from two data sets $y$ and $y^\prime$ under the assumption of identical priors $\pi(\theta)$ leads to
\begin{equation}
S(y,y^\prime) =
\int\dd\theta\:p(\theta|y)\ln\left(\frac{p(\theta|y)}{p(\theta|y^\prime)}\right) =
\int\dd\theta\:p(\theta|y)\ln\left(\frac{\likeli(y|\theta)}{\likeli(y^\prime|\theta)}\frac{p(y^\prime)}{p(y)}\right) =
\int\dd\theta\:p(\theta|y)\ln r - \ln B \,,
\label{eq:KL_data}
\end{equation}
because the posterior distribution is normalised, $\int\dd\theta\:p(\theta|y) = 1$. In Eq.~(\ref{eq:KL_data}) one can identify the Bayes-ratio $B = p(y)/p(y^\prime)$ of evidences and the likelihood ratio $r = \likeli(y|\theta)/\likeli(y^\prime|\theta)$. Typically, the Bayes-ratio is used to compare different models, as the evidence differs for different models, but equal data. Here, the model remains the same, but two different data sets are considered, generating the deviation from unity in the Bayes-ratio. The relative entropy $S(y,y^\prime)$ can be interpreted as a measure of compatibility of the posterior distributions derived from two different data sets and consists of the two contributions: The likelihood ratio $r$ averaged over posterior distribution and the Bayes-ratio $B$: The two posterior distributions surely differ according of how probable the observation of the respective data set was under the assumption of the parameter choice, but the second contribution suggests the influence of the posterior, too: It would matter to $S(y,y^\prime)$ how probable the observation of $y$ and $y^\prime$ could have been given the prior \citep{2006PhRvD..73f7302M, 2002MNRAS.335..377H, 2015MNRAS.449.2405K, 2019JCAP...01..036A}: In general, different data sets would generate different Bayesian evidences, which by itself is a sensible (and sensitive) discriminant. In some sense, Eq.~(\ref{eq:KL_data}) is remarkable as it combines the likelihood ratio $r$ as a typical frequentist quantity with Bayes-ratio $B$ in a single formula in this way. 

It is possible to define a metric distance for every conditional probability, as motivated in the introduction. Then, the Fisher metric $F_{ij}$ derived from the posterior $p(\theta|y)$ for a data set $y$ and an infinitesimally different data set $y + \dd y$ would lead to
\begin{equation}
S(y,y+\dd y) =
\int\dd\theta\:p(\theta|y)\ln\left(\frac{p(\theta|y)}{p(\theta|y+\dd y)}\right) =
\frac{1}{2}F(y)_{ij}\dd y^i\dd y^j
\label{eq:KL_fisher_approx}
\end{equation}
and would have the interpretation of data consistency between two infinitesimally distant data sets with respect to implying similar posterior distributions. For the particular choice of Kullback-Leibler divergences one can derive the Fisher metric $F_{ij}$ from $ S(y,y+\dd y)$,
\begin{equation}
F_{ij}(y) = \int\dd\theta\:p(\theta|y)\:\partial_i\ln p(\theta|y)\partial_j\ln p(\theta|y) \,,
\label{eq:fisher_def}
\end{equation}
with derivatives $\partial_i = \partial/\partial y^i$. Then, finite differences $\Delta y^i$ between two data sets $y$ and $y^\prime = y + \Delta y$ defines a metric line element $s^2_y$
\begin{equation}
S(y,y^\prime) \simeq \frac{1}{2}F(y)_{ij}\Delta y^i\Delta y^j = \frac{1}{2}s^2_y \,,
\label{eq:data_distance}
\end{equation}
under the assumption of a constant Fisher metric $F(y)_{ij}$. Typically, one has Gaussian error processes for the data $y$ with $S(y,y+\Delta y) = \Delta\chi^2/2$, such that $\Delta\chi^2 = F_{ij}\Delta y^i\Delta y^j = s^2_y$, agreeing with expectations. Due to the choice of natural logarithms, both $S$ and $s_y^2$ are measured in nats.

Interestingly, the interpretation of the conventional Kullback-Leibler divergence in parameter space between likelihoods $\likeli(y|\theta)$ and $\likeli(y|\thetap)$ as the starting point of the derivation of the Fisher metric $F_{\mu\nu}(\theta)$ is not as clear: Repeating the same steps of inserting Bayes' law into the Kullback-Leibler measure for the same data set $y$, identical evidence $p(y)$ implies
\begin{equation}
S(\theta,\thetap) =
\int\dd y\:\likeli(y|\theta)\ln\left(\frac{\likeli(y|\theta)}{\likeli(y|\thetap)}\right) =
\int\dd y\:\likeli(y|\theta)\ln\left(\frac{p(\theta|y)}{p(\thetap|y)}\frac{\pi(\thetap)}{\pi(\theta)}\right) =
\int\dd y\:\likeli(y|\theta)\ln\left(\frac{p(\theta|y)}{p(\thetap|y)}\right) + \ln\frac{\pi(\thetap)}{\pi(\theta)} \,,
\end{equation}
because the likelihood is normalised, $\int\dd y\:\likeli(y|\theta) = 1$. The relation implies that the relative entropy, which is to become a metric distance in the framework of information geometry, has two contributions, one from the likelihood-weighted logarithmic ratio between the posteriors at given $\theta$ and $\theta^\prime$, as well as from the logarithmic prior ratio.

Much clearer, fortunately, is the interpretation of the conventional Fisher metric $F_{\mu\nu}$ derived from the likelihood $\likeli(y|\theta)$ for a parameter choice $\theta$ and an infinitesimally distant parameter choice $\theta+\dd\theta$, for fixed data $y$
\begin{equation}
S(\theta,\theta+\dd\theta) =
\int\dd y\:\likeli(y|\theta)\ln\left(\frac{\likeli(y|\theta)}{\likeli(y|\theta+\dd\theta)}\right) = \frac{1}{2} F(\theta)_{\mu\nu}\dd\theta^\mu\dd\theta^\nu
\end{equation}
as the difference in goodness of fit, which becomes a line element $\dd\chi^2$ in parameter space. With derivatives $\partial_\mu = \partial/\partial\theta^\mu$ one arrives at \citep{amari_information_2016}
\begin{equation}
F_{\mu\nu}(\theta) =
\int\dd y\:\likeli(y|\theta)\:\partial_\mu\ln\likeli(y|\theta)\partial_\nu\ln\likeli(y|\theta) \,,
\end{equation}
such that two model choices $\theta$ and $\thetap = \theta + \Delta\theta$ with finite differences have a metric distance
\begin{equation}
S(\theta,\thetap) \simeq \frac{1}{2}F_{\mu\nu}(\theta)\Delta\theta^\mu\Delta\theta^\nu = \frac{1}{2}s^2_\theta
\end{equation}
approximating the metric distance $s^2_\theta$ by assuming a constant $F_{\mu\nu}(\theta)$: Otherwise, the computation of an arc length would be required in a line integral along a parameterised curve $\theta^\mu(\lambda)$ \citep{giesel2020information}.

\subsection{Bayes' theorem from the point of view of information entropy}
As the Fisher-matrices $F_{ij}$ and $F_{\mu\nu}$ are derived from the posterior distribution $p(\theta|y)$ and the likelihood $\likeli(y|\theta)$, respectively, there should be a deeper relation between these relative entropies and the Bayes' theorem \citep{baez_bayesian_2014,1056144}: Starting with the mutual entropy for the joint probability $p(\theta,y) = p(\theta|y)p(y) = \likeli(y|\theta)\pi(\theta)$, and with $\pi(\theta)$ and $p(y)$ are the marginals of the joint distribution $p(\theta,y)$, one defines the conditionals: $\likeli(y|\theta) = p(y,\theta)/\pi(\theta)$ and $p(\theta|y) = p(y,\theta)/p(y)$. Then, the mutual entropy $S(\theta,y)$ is given by
\begin{equation}
S(\theta,y) =
\int\dd\theta\:\int\dd y\: p(\theta,y)\ln\frac{p(\theta,y)}{\pi(\theta)p(y)} =
\int\dd y\: p(y)\:\int\dd\theta\: p(\theta|y)\ln\frac{p(\theta|y)}{\pi(\theta)} \,,
\end{equation}
and would vanish if $p(\theta,y)$ factorizes into $\pi(\theta)p(y)$, which can only be the case if the data contains no information about the physical model. In general, though, the evidence-weighted Kullback-Leibler divergence between posterior and prior distributions corresponds to the surprise statistic \citep{Seehars:2014ora,mello2024nongaussiansurprisestatisticcosmological}, see also \citep{2006PhRvD..74b3503K}.

Vice versa, one obtains
\begin{equation}
S(y,\theta) =
\int\dd\theta\:\int\dd y\: p(\theta,y)\ln\frac{p(\theta,y)}{\pi(\theta)p(y)} =
\int\dd\theta\:\pi(\theta)\:\int\dd y\:\likeli(y|\theta) \ln \frac{\likeli(y|\theta)}{p(y)} \,,
\end{equation}
implying that the surprise statistic is given by the prior-averaged information entropy of the likelihood minus the information entropy of the posterior distribution:
\begin{equation}
\left\bra S(\pi(\theta)\rightarrow p(\theta|y))\right\ket_{p(y)} =
\left\bra S(p(y)\rightarrow \likeli(y|\theta)\right\ket_{\pi(\theta)} \,,
\label{eqn_surprise}
\end{equation}
pointing at an interpretation of the surprise statistic as the relative entropy between likelihood and evidence, averaged over the prior allowed parameter range.

\section{Fisher metric in data space}\label{sect_fisher}
The Fisher metric $F_{ij}(y)$ in data space quantifies the difference between the posterior distributions derived from two infinitesimally different data sets. In an application to a cosmological inference problem, for instance, the tension between the values for the Hubble-Lema{\^i}tre constant $H_0$ from supernovae and CMB-measurements, one would proceed like this: The first data set implies a posterior distribution from which one can construct a hypothetical distribution of data sets of the second measurement. Then, one can compare this hypothetical second measurement to the actual second measurement and quantify the amount of compatibility with a Kullback-Leibler-type measure, or in approximation, with a distance induced by the Fisher metric $F_{ij}(y)$. It is important to realise that in this way $(i)$ one compares the compatibility of two data sets, allowing $(ii)$ a ranking of data points in their influence on the posterior distribution. And $(iii)$, it would be the case as for all relative entropies \citep[with the exception of the Bhattacharyya-entropy,][]{Bhattacharyya_original} that the distance measure is not symmetric: In Eq.~(\ref{eq:fisher_def}) it matters which data set is used as the basis for the posterior distribution $p(\theta|y)$.

The Fisher metric can be rewritten by using 
\begin{equation}
\langle \partial_i\ln p(\theta|y) \partial_j\ln p(\theta|y) \rangle =
-\langle \partial_i \partial_j\ln p(\theta|y) \rangle \, ,
\end{equation}
with the expectation value $\langle \, . \, \rangle$ computed by integration over the random variable $\theta$. Then, Eq.~(\ref{eq:fisher_def}) becomes
\begin{equation}
F_{ij} = - \int\dd\theta\:p(\theta|y)\: \partial_i \partial_j\ln p(\theta|y)\, .
\end{equation}
To calculate the partial derivatives we can use Bayes' theorem Eq.~(\ref{eq:bayes_theorem}) and substitute the posterior, yielding:
\begin{equation}
\partial_i \partial_j\ln p(\theta|y) =
\partial_i \partial_j \ln\left(\frac{\likeli(y|\theta)\pi(\theta)}{p(y)}\right) =
\partial_i \partial_j \left(\ln \likeli(y|\theta)-\ln p(y)\right)\,,
\end{equation}
where the derivative of the prior vanishes, as it does not depend on the data.

To continue with our calculation we need to assume the explicit shape of the likelihood. Here, we will assume, that we have a Gaussian likelihood with inverse covariance matrix $\Sigma_{ij}$, data points $y^i$ and a model prediction $y(\theta)$, resulting in the following shape:
\begin{equation}
\likeli \propto \exp\left\{-\frac{1}{2} ( y-y(\theta))^i \Sigma_{ij}( y-y(\theta))^j\right\}
\label{eq:likelihood_form}
\end{equation}
Note, that the likelihood is only Gaussian in the data and not in the parameters, as $y(\theta)$ can be any function. Now, the derivatives of the log-likelihood follow as
\begin{equation}
\partial_i \partial_j \ln \likeli(y|\theta) =
-\partial_i \partial_j \frac{1}{2}\sum_k\left( ( y-y(\theta))^l \Sigma_{lk}( y-y(\theta))^k\right) =
-\partial_i \left(\Sigma_{jk}( y-y(\theta))^k\right) =
-\Sigma_{ij}\,,
\end{equation}
with $\partial_i \Sigma_{jk}=0$, due to the small impact of varying data on the covariance matrix \citep{schafer_describing_2016,Reischke:2016ana}.
The derivative of the evidence is more involved. We will start by explicitly performing the derivative and substituting the formula for the evidence Eq.~(\ref{eq:evidence}):
\begin{align}
- \partial_i \partial_j \ln p(y)&= \frac{1}{p(y)^2} \partial_i p(y) \partial_j p(y) -\frac{1}{p(y)} \partial_i \partial_j p(y) \\
&= \frac{1}{p(y)^2} \int \pi(\theta)\: \partial_i \likeli(y|\theta) \dd \theta \, \int \pi(\theta)\: \partial_j \likeli(y|\theta) \dd \theta - \frac{1}{p(y)} \int \pi(\theta)\: \partial_i \partial_j \likeli(y|\theta) \dd \theta \, ,
\end{align}
where we used the interchangeability of integral and derivative and that the prior does not depend on the data. Now, the first and second partial derivatives of the likelihood are
\begin{equation}
\partial_i \partial_j \likeli(y|\theta) =
-\partial_i \left(\likeli(y|\theta)\Sigma_{jk}( y-y(\theta))^k\right) =
\likeli(y|\theta)\Sigma_{il}( y-y(\theta))^l \Sigma_{jk}( y-y(\theta))^k - \likeli(y|\theta) \Sigma_{ij}\, .
\end{equation}
Putting it all together the derivatives of the log posterior read as
\begin{align}
\partial_i \partial_j\ln p(\theta|y) = -\Sigma_{ij} + \int  \frac{\likeli(y|\theta)\pi(\theta)}{p(y)} \Sigma_{il}( y-y(\theta))^l \dd \theta \, \int  \frac{\likeli(y|\theta)\pi(\theta)}{p(y)} \Sigma_{jk}( y-y(\theta))^k \dd \theta \nonumber\\
 - \int \frac{\likeli(y|\theta)\pi(\theta)}{p(y)}\Sigma_{il}( y-y(\theta))^l\Sigma_{jk}( y-y(\theta))^k \dd \theta + \int  \frac{\likeli(y|\theta)\pi(\theta)}{p(y)} \Sigma_{ij} \dd \theta\, ,
\end{align}
where we can identify Bayes' theorem multiple times and simplify it to
\begin{align}
\partial_i \partial_j\ln p(\theta|y) = \Sigma_{il} \Sigma_{jk}\left(\int p(\theta|y) ( y-y(\theta))^l \dd \theta  \, \int p(\theta|y) ( y-y(\theta))^k \dd \theta - \int p(\theta|y) ( y-y(\theta))^l( y-y(\theta))^k \dd \theta \right)\, .
\end{align}
Here, the terms only with $ \Sigma_{ij}$ cancel each other and the rest are expectation values with respect to the posterior:
\begin{align}
\partial_i \partial_j\ln p(\theta|y) &= \Sigma_{il} \Sigma_{jk}\left(\langle (y-y(\theta))^l \rangle_{p(\theta|y)} \langle (y-y(\theta))^k \rangle_{p(\theta|y)}  - \langle (y-y(\theta))^l (y-y(\theta))^k \rangle_{p(\theta|y)}\right)\nonumber\\
&= \Sigma_{il} \Sigma_{jk}\left(\langle y^l(\theta)\rangle_{p(\theta|y)} \langle y^k(\theta)\rangle_{p(\theta|y)} - \langle y^l(\theta)y^k(\theta)\rangle_{p(\theta|y)}\right)\, ,
\label{eq:deriv_posterior}
\end{align}
where the last simplification was obtained by collecting terms. It is important to note, that Eq.~(\ref{eq:deriv_posterior}) does not depend on the model parameters $\theta$ as they are integrated out by taking the expectation value. Therefore, the Fisher metric is given by
\begin{equation}
F_{ij} = - \Sigma_{il} \Sigma_{jk}\left(\langle y^l(\theta)\rangle_{p(\theta|y)} \langle y^k(\theta)\rangle_{p(\theta|y)} - \langle y^l(\theta)y^k(\theta)\rangle_{p(\theta|y)}\right)\, ,
\end{equation}
which is a product of the inverse covariance matrices of the data points times the covariance matrix of the model predictions:
\begin{equation}
F_{ij} = \Sigma_{il} \Sigma_{jk} \mathrm{cov}_{p(\theta|y)}[y^l(\theta),y^k(\theta)]
\label{eq:fisher_final}
\end{equation}

\section{Application to the cosmological tension between supernova and CMB data}\label{sect_cosmo}
The tension in the Hubble-Lema{\^i}tre constant $H_0$ arising from the two measurements, supernovae type Ia (SNIa) with cepheids and the cosmic microwave background (CMB), is typically measured in $\sigma$. The formulation of the Fisher metric in data space allows us to describe the distance between datasets in a natural way. The goal of this section is to obtain a metric on the data space of the SNIa measurements as well as the data space of the CMB spectra and obtain a distance between SNIa data and CMB data. In the following we will abbreviate the combination of cepheids and supernova with supernova.

\subsection{Likelihoods}\label{sec:like}
For this analysis, we need likelihoods for the supernovae data and the CMB data, since samples from the posterior are necessary to calculate the data space fisher metric defined in Eq.~(\ref{eq:fisher_final}). This section briefly summarises the likelihood and data for the SN1a and the CMB.

\subsubsection{Supernova Likelihood}
The distance modulus is defined through the total and apparent magnitude as
\begin{align}
\mu = m-M = 5\log_{10}\frac{d_L}{10\text{pc}}\, ,
\label{eq:distance_modulus}
\end{align}
where $d_L$ is the luminosity distance. In this analysis, the unit of $d_L$ is kpc, which allows to rewrite Eq.~(\ref{eq:distance_modulus}) as
\begin{align}
\mu = 5\log_{10}d_L+10\, .
\end{align}
The luminosity distance is defined via the differential equation
\begin{align}
\frac{\dd d_L}{\dd z} = \frac{d_L}{1+z}+c \frac{1+z}{H(z)}\, ,
\end{align}
with
\begin{align}
H^2(z) = H_0^2\left(\Omega_\text{m}(1+z)^3+\Omega_w(1+z)^{3(1+w)}\right)\,,
\end{align}
where $H_0=h\times 100\,\text{(km/s)/Mpc}$ and $c$ is the speed of light. For simplicity of the analysis, we constrain ourselves to a spatially flat $\Lambda$CDM-universe, which means setting $w=-1$. $\Omega_w$ can simply be calculated with $\Omega_\Lambda = 1 - \Omega_\text{m}$. Therefore, given the parameter set $\theta=\{\Omega_\text{m},h\}$ one can calculate the distance modulus as a function of redshift $z$, denoted by $\mu(z|\theta)$. \\
The SNIa dataset is Pantheon+SH0ES \citep{Scolnic_2022,Yuan_2022}, with 1701 datapoints of which 77 are anchored. We use the likelihood defined in \cite{Brout_2022}. Due to the degeneracy of $h$ and $M$ one needs distance anchors, which are provided by Cepheid calibrated host-galaxies. 
The likelihood reads
\begin{align}
\likeli(m|\theta) \propto \exp\left\{-\frac{1}{2}  \Delta D^i \Sigma_{ij}\Delta D^j\right\}\,,
\label{eq:likelihood_sn}
\end{align}
with 
\begin{align}
    \Delta D^i = 
    \begin{cases}
        \mu^i - \mu^i_\text{Cepheid} & i \in \text{Cepheid hosts}\\
        \mu^i - \mu^i(z|\theta) & \text{otherwise}
    \end{cases}\,,
\end{align}
where $m^i = \mu^i+M$ and $\Sigma_{ij}$ are the measurements and inverse covariance matrix for the apparent magnitude and $z$ is the redshift of the respective measurement. The parameter set is $\theta=\{\Omega_\text{m},h,M\}$, where we marginalise over $M$ to obtain the same parameter set as in the CMB analysis. The likelihood has the same general form as is required in Eq.~(\ref{eq:likelihood_form}).

\subsubsection{CMB Likelihood}
For the analysis of the CMB data, we use the Planck-lite high-$\ell$ likelihood, presented in \citep{2020_planck_like}. Compared to the full Plik likelihood, all except one nuisance parameter are integrated out. This saves us a lot of computational cost at a marginal loss in precision, which is not relevant to this conceptual work. We consider base $\Lambda \text{CDM}$ parametrized by the parameter set $\{\Omega_m,\,h,\,\omega_\text{b},\,\ln10^{10}A_s,\,n_s,\,\tau_\text{reio}\}$. Over all parameters, except $\Omega_m$ and $h$ is marginalized to obtain a comparable posterior.

The $C_{\ell}$ are averaged into 215 bins, where $\ell_\text{min}=30$ and $ \ell_\text{max}=2508$. The spectra are calculated with the  Cosmic Linear Anisotropy Solving System (CLASS) \citep{CLASS_lesgourges}. The likelihood is given by
\begin{equation}
    \likeli(C_\ell|\theta) \propto \exp\left\{-\frac{1}{2}(C_\ell-C_\ell(\theta))^i\Sigma_{ij}(C_\ell-C_\ell(\theta))^j\right\}\,,
\end{equation}
with the inverse covariance matrix $\Sigma_{ij}$, the data $C_\ell$ and the spectra calculated from given parameters $C_\ell(\theta)$. Note, that we have the same functional shape as in Eq.~(\ref{eq:likelihood_sn}). The posterior samples are obtained with an MCMC analysis using MontePython \citep{Brinckmann:2018cvx}.

\begin{figure}[t!]
    \centering
    \includegraphics[width=\columnwidth]{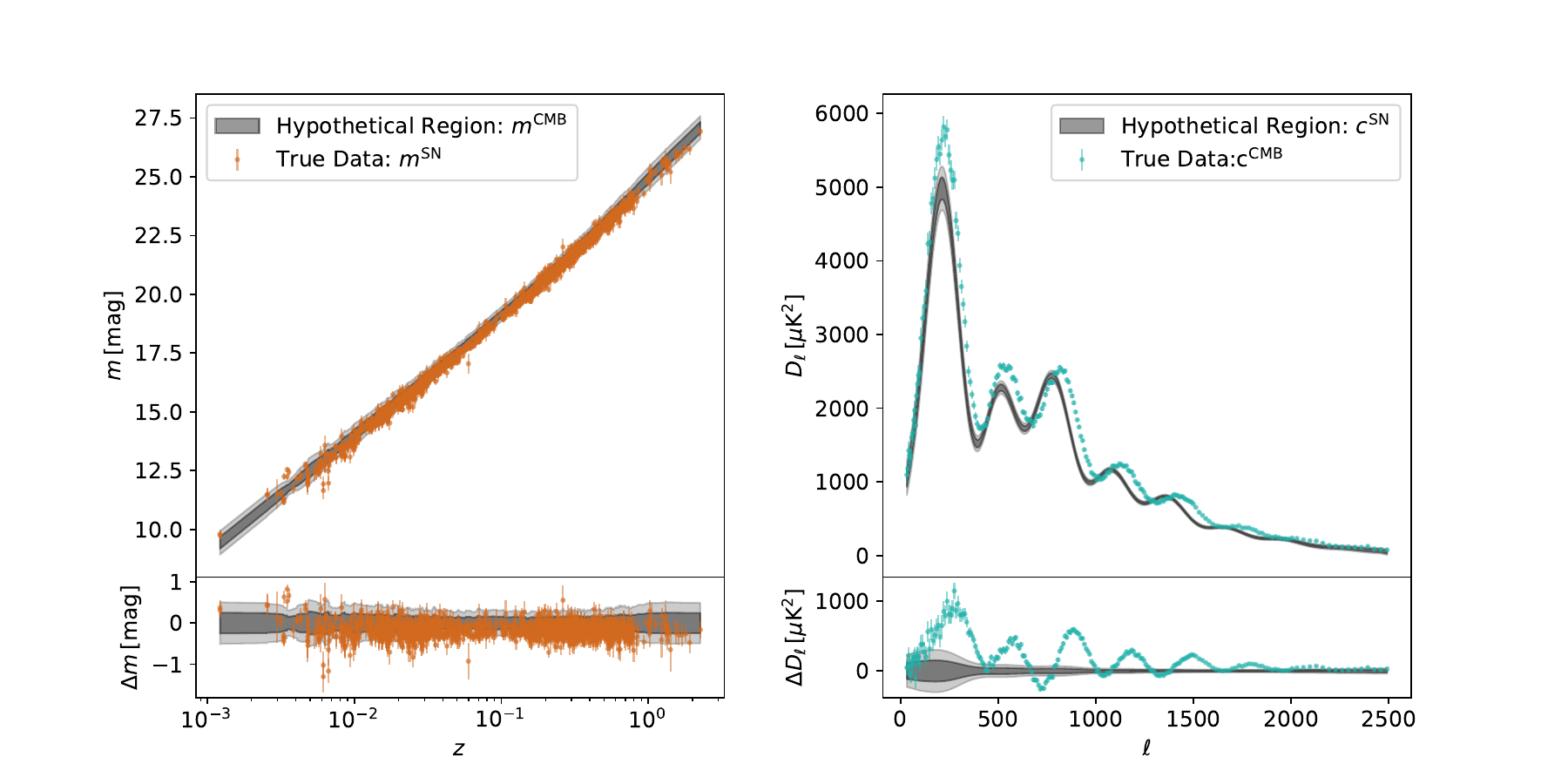}
    \caption{True data from the Panthone+SH0ES supernova dataset (orange) and the Planck CMB power spectrum (blue). Translating the inferred best-fit parameters from one experiment into hypothetical expected data from the other experiment results in the grey-shaded region.}
    \label{fig:data_vis}
\end{figure}

\begin{figure}[t]
    \centering
    \includegraphics[width=.39\columnwidth]{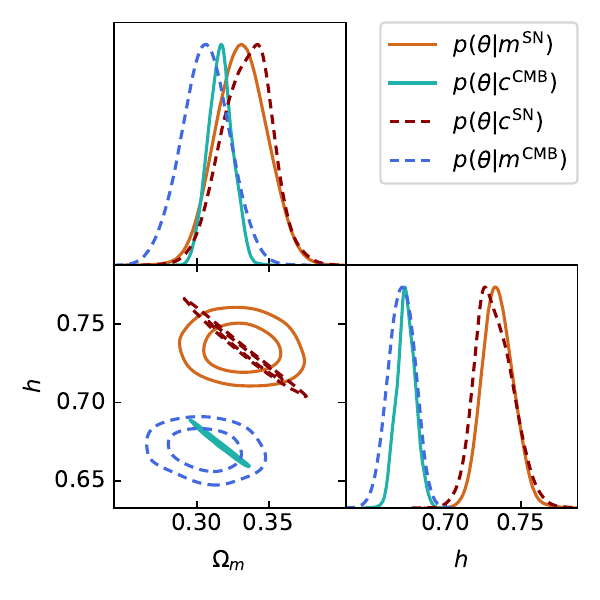}
    \caption{Inferred 2d and 1d marginalized posteriors from supernova and CMB data, shown in Fig.~\ref{fig:data_vis}. The four posteriors emerge from considering real and hypothetical data, the grey regions in Fig.~\ref{fig:data_vis}. The dotted posteriors are from hypothetical data. Note the coinciding maxima of the posteriors when the data has the superscript CMB or SN, respectively, and the same width if the data comes from the same data space, $\mu$ or $c$.}
    \label{fig:posterior}
\end{figure}

\subsection{Translation between data spaces}\label{sec:translation}
The measurements obtained from Cosmic Microwave Background (CMB) and Type Ia Supernovae (SNIa) experiments are characterized by different data spaces. In the context of CMB, the spectral information is represented by $C_\ell$, while SNIa measurements are expressed through the apparent magnitude $m$. The necessity to compare and combine these datasets, as required by Eq.~(\ref{eq:fisher_final}), mandates the translation of data into a unified data space. Given the absence of a preferred data space, both CMB and SNIa data spaces are examined.

To facilitate the translation between these spaces, two models are introduced: $\mathcal{M}^\text{SN}$ and $\mathcal{M}^\text{CMB}$. These models enable the computation of data for both spaces using the same underlying parameters. Explicitly, they are both $\Lambda\text{CDM}$, but for different times of the universe. Consequently, starting with a real dataset $c^\text{CMB}$, one can deduce the best-fit parameters $\theta^\text{CMB}$ and generate hypothetical data in the SN space through $\mathcal{M}^\text{SN}(\theta^\text{CMB})=m^\text{CMB}$. The concept of hypothetical data corresponding to different parameter values is illustrated in Fig.~\ref{fig:data_vis}. By using this dataset $\mu^\text{CMB}$ to infer best-fit parameters, one would recover $\theta^\text{CMB}$.

Since the uncertainties of data points are influenced by the statistics and systematics of the original experiment, it is assumed that $m^\text{CMB}$ and the real measurement $m^\text{SN}$ share the same covariance matrix. This results in distinct posteriors for the two data spaces, specifically $p(\theta|m^\text{CMB})\neq p(\theta|c^\text{CMB})$. The visual representation of these four posteriors, two based on real data and two on hypothetical data, is illustrated in Fig.~\ref{fig:posterior}. The tension between the real data becomes evident, as the contours for $p(\theta|c^\text{CMB})$ and $p(\theta|m^\text{SN})$ do not overlap. The posteriors derived from hypothetical data successfully recover the means, but exhibit different widths in their distributions, highlighting the respective experiment's ability to constrain the chosen parameter set.

\subsection{Distances}\label{sec:distance}
The previous Sect.~\ref{sec:translation} showed how we can compare data from two different data spaces. Now, we can compare SNIa and CMB data in the two possible spaces. The measure of compatibility is the distance between the two datasets, calculated with Eq.~(\ref{eq:KL_fisher_approx}), i.e. $s(m^\text{SN},m^\text{CMB})$ and  $s(c^\text{SN},c^\text{CMB})$, where the previously used $y$ is now $m$ or $c$. For the rest of this work, $y \in \{m,c\}$.

Intrinsically, $s_y$ does not have an error, however, the data does. We expect the data to be on the best-fit cosmology line, yet data is distributed according to a Gaussian distribution. Therefore, it is sensible to resample the hypothetical data from this distribution and recalculate $s_y$ to estimate the uncertainty. We observe, that the the Fisher metric is not influenced by the resampling. This is to be expected because the posterior is unaffected and $F$ only depends on its shape. The effect of the resampling is solely in $\Delta y = y^\text{SN} - y^\text{CMB}$.

\begin{figure}[tp!]
    \centering
    \includegraphics[width=.68\columnwidth]{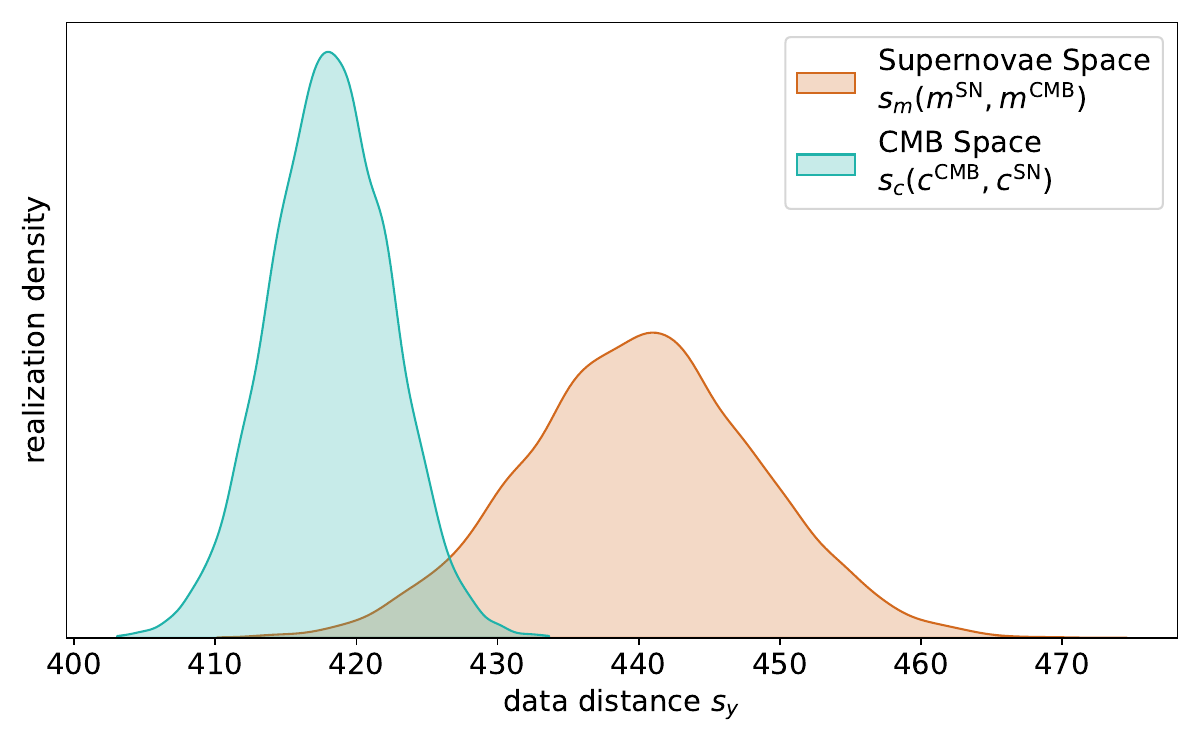}
    \caption{Distribution of the data distance (\ref{eq:data_distance}) between the CMB and supernova dataset. The left panel (orange) shows the distance in SN space for many realizations of hypothetical data, corresponding to CMB best-fit parameters. The right panel (blue) displays the corresponding scenario in the CMB space with realizations of hypothetical data calculated from SN best-fit parameters. }
    \label{fig:distance}
\end{figure}

The resulting distribution for $s_y$ is shown in Fig.~\ref{fig:distance}. The values read $s^\text{SN}_m= 440\pm 8$ and $s^\text{CMB}_c = 418\pm 4$. Note, that we show an actual metric distance, which is related to the KL-divergence with $ S =  s_y^2/2$, whose value is to be interpreted in nats. The values are very similar in both data spaces and show a tension. 

Having calculated the Fisher metric allows us to further investigate the influence of individual data points on the tension. We can investigate which data point has the largest impact on the distance, by calculating the logarithmic derivative
\begin{align}
\frac{\partial s_y}{\partial \ln y^i} = 
\frac{\sum_j F_{ij} y^i \Delta y^j}{s_y}\,,
\end{align}
where the Einstein summation convention is not applied and we explicitely write the sum over $j$. This derivative combines the error of the data point with its actual value in order to quantify the leverage of $y^i$ for a fixed index $i$ on the posterior distribution in a dimensionless way.

\begin{figure}[tp!]
    \centering
    \includegraphics[width=\columnwidth]{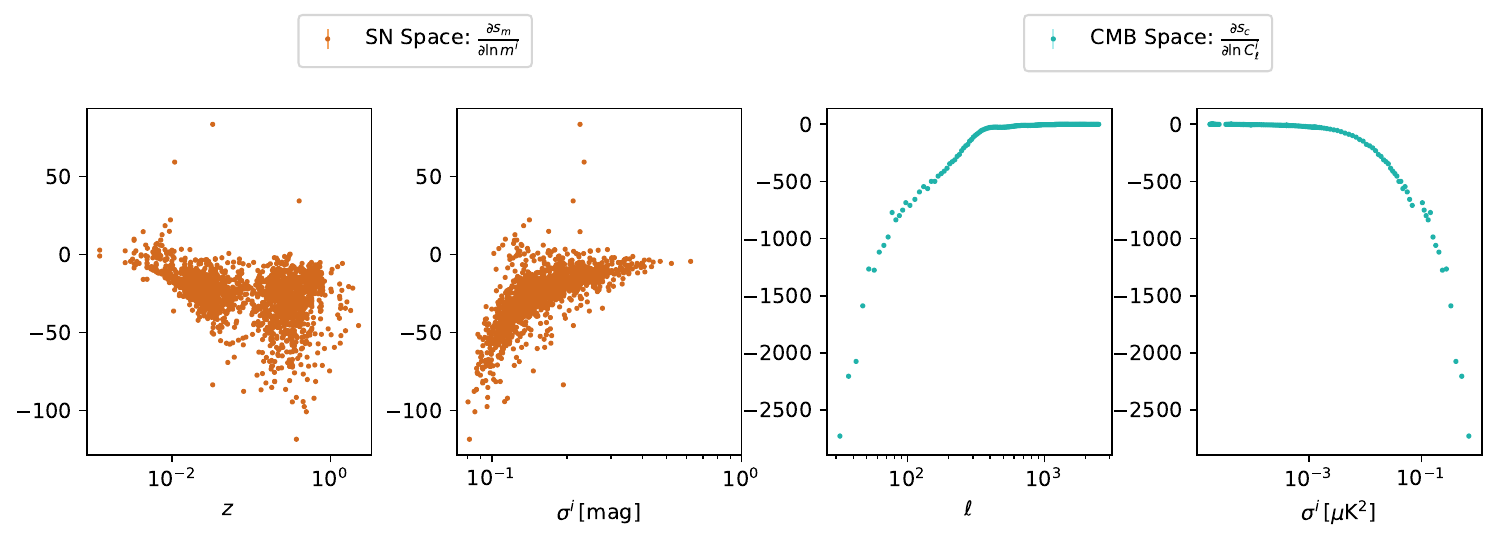}
    \caption{Importance of each data point quantified by the log derivative of distance. Error erstimates through resampling included but to small to be visible. The left side (orange) is in the SN space and the right (blue) is in the CMB space. The first panel shows which redshifts are most important. The third panel shows the same concept but for multipole moments in the CMB space. The second and fourth panels show how the uncertainty of the individual measurement influences the capability to reduce the distance in the SN and CMB space, respectively.}
    \label{fig:derivative}
\end{figure}

The idea of this object is to quantify the possible impact of one data point on the distance between two datasets. Therefore, characterising which data points are most important for a precise measurement. The uncertainty on this object is calculated by resampling the possible realizations of the hypothetical data. The results are shown in Fig.~\ref{fig:derivative}.

The first panel relates the derivate of the distance with the redshift of the data point. Almost all points show a negative derivative, implying that the distance between the two datasets reduces if the measured supernova would be further away, which is consistent with what one would expect intuitively. There is no clear visible trend that would point towards which redshifts are most important. There are some outliers with the largest impact on the distance, allowing us to label them as the most important data. The second panel relates the derivate with the uncertainty of the measurement. Here, a small uncertainty corresponds to a large impact on the distance and vice versa.

The third and fourth panels show the same as the first and second, but now for the CMB space. Low-$\ell$ multipoles have a much larger pull on the distance than high-$\ell$, even though their uncertainty is larger. This can be observed in the fourth panel, where we see the inverse relation, compared to the SN scenario. The uncertainty of the low-$\ell$ region is dominated by cosmic variance and a more precise measurement is not feasible. Therefore, more accuracy up to $\ell \approx 500 $ shows the largest potential, either in theory or in systematics.

\begin{figure}[tp!]
    \centering
    \includegraphics[width=.7\columnwidth]{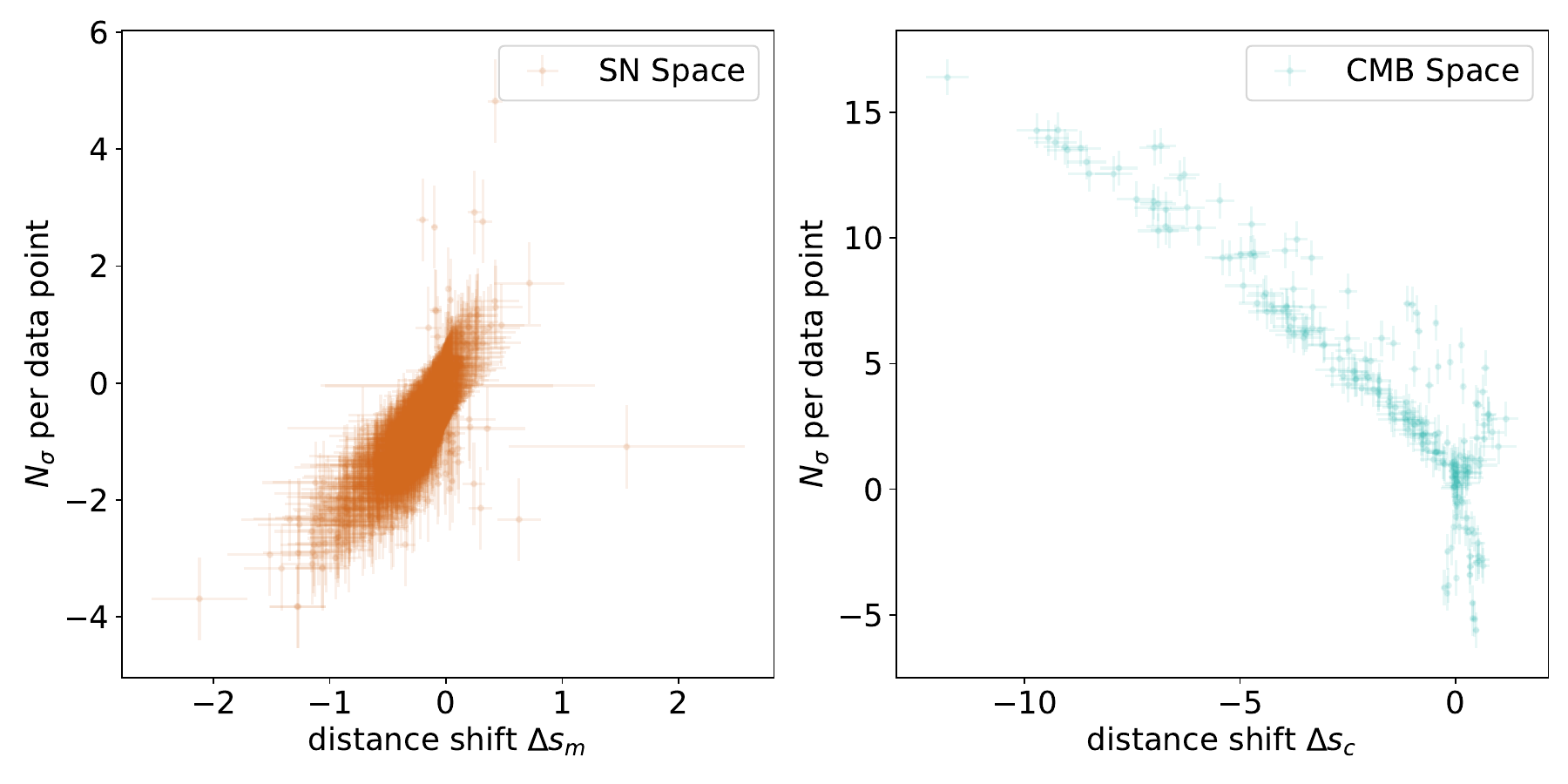}
    \caption{Relation between $N_\sigma$, the number of standard deviations, one hypothetical and real data point differ, and the shift in distance induced by changing one of the true data points to the hypothetical one. The left panel (orange) shows the SN space and the right (blue) the CMB space.}
    \label{fig:distance_shifts}
\end{figure}

Coming back to the tension between the two datasets, we can test which data point has the largest influence on the incompatibility if it were to be measured incorrectly. Here, we assume that all of them are independent: While this is technically not the case due to correlations in calibration on couplings between binned multipoles, these correlations are weak and could nevertheless be incorporated into the analysis. The procedure is to propose a hypothetical dataset, where the $i$th point of one dataset is altered, such that it is equal to the $i$th point in the other set, i.e. $y^{i,\text{SN}}=y^{i,\text{CMB}}$, where the $i$ is fixed for each recalculation. We change the value of the true datasets and compare it to the hypothetical ones. We want to compare this idea of quantifying the influence of a wrong measurement with the number of sigmas between two data points $N_\sigma$. Here, $N_\sigma$ is allowed to be negative to indicate whether a data point is too high or too low, as shown by Fig.~\ref{fig:distance_shifts}. There is a trend of proportionality, even more pronounced in the CMB space. In the CMB plot, there are two main lines, which come from the oscillating behaviour of the spectrum. In the SN space, multiple points exist that differ up to $2\sigma$, but do not shift the distance. Additionaly, there is data which would increase the distance if they were measured at the assumed value. 

\section{summary and discussion}\label{sect_summary}
In our study we apply methods from information geometry to the issue of cosmological tensions, in particular, the tension between estimates of the Hubble-Lema{\^i}tre constant $H_0$ on the basis of supernova and CMB data, which is highly significant in Gaussian terms, and widely discussed in the literature as a signature of new physics beyond $\Lambda$CDM.

\begin{enumerate}[(i)]
\item{In terms of information geometry, tensions or dissimilarities between conditional distributions are quantified by axiomatically defined statistical divergences and suggest a metric distance in the manifold of their parameters. We use the symmetry of Bayes' law for establishing Kullback-Leibler divergences for the posterior distributions in their dependence on data, which becomes the basis of an information geometrical distance measure. In particular, this relative information entropy seems to be composed of the averaged likelihood ratio and the Bayesian evidence ratio as an intuitive explanation. As entropies are measured in units of nats, the distance measure on the data manifold would correspondingly be quantified in $\sqrt{\mathrm{nats}}$. In contrast to intuitive measures like the Mahalanobis distance, the information theoretical distance does not rely on Gaussianity of the distributions.}
\item{There seems to be a relation between the surprise statistic as the Kullback-Leibler divergence between the prior and the posterior distribution. Starting from their mutual information and by employing Bayes' law one can show that the surprise statistic corresponds to relative entropy between the likelihood and the Bayes evidence as distributions of the expected data.}
\item{We present an explicit derivation of the Fisher metric $F_{ij}$ in data for a Gaussian error process in the measurement. The result is applied to the tension in estimates for the Hubble-Lema{\^i}tre constant $H_0$ from CMB and supernova measurements. For this purpose, we ask what supernova apparent magnitudes one would have expected in the light of the CMB posterior, and what CMB spectra one would have observed if the supernova posterior had been true, with a propagation of the uncertainty contained in the prior of the first data set onto an error band for the second data set, with a subsequent estimation of the information theoretical distance measure. The origin of this tension is generated by supernovae in the entire redshift range if the CMB data is considered to be correct. But vice versa, it is predominantly the low-$\ell$ CMB spectrum that is responsible for the tension if the supernova posterior is valid. We would like to emphasise, though, that statistical divergences like Kullback-Leibler do not differentiate between the relative centroid shift of two distributions and their differing width.}
\item{Using the data space Fisher metric as a diagnostic tool, it is possible to disentangle the two contributions to the power of a data point to influence the posterior distribution: Clearly, an outlying data point can influence the fit according to its distance from the fit suggested by the other data points, but only to the extend how strongly the model varies at that particular point. Our study seems to suggest that in the case of supernova data there, almost all data points contribute to the tension irrespective of their significance and that in the case of the CMB data, the significant deviations in the low-$\ell$ data points predominantly generate the tension.}
\end{enumerate}

It would be interesting to see if Eq.~(\ref{eqn_surprise}) could provide means by detecting anomalies in data, developing the idea of this paper further: In an abstract way, one could quantify the surprise statistic in data space for individual data points and quantify outliers using this methodology.

\section*{Acknowledgements}
We would like to thank Alan Heavens for an insightful conversation.

\paragraph{Funding information}
BS and BMS would like to acknowledge financial support by the Vector Stiftung. We acknowledge the usage of the AI-clusters {\em Tom} and {\em Jerry} funded by the Field of Focus 2 of Heidelberg University. MQ is supported by the Brazilian research agencies FAPERJ, CNPq (Conselho Nacional de Desenvolvimento Científico e Tecnológico) and CAPES. This study was financed in part by the Coordenação de Aperfeiçoamento de Pessoal de Nível Superior - Brasil (CAPES) - Finance Code 001. We acknowledge support from the CAPES-DAAD bilateral project  ``Data Analysis and Model Testing in the Era of Precision Cosmology''.

\paragraph{data availability statement}
No new data were generated or analysed in support of this research.

\vspace{1cm}

\bibliographystyle{mnras}
\bibliography{references}

\end{document}